\documentclass[aps,prl,twocolumn,showpacs,superscriptaddress,groupedaddress]{revtex4}  

\usepackage{amssymb}
\usepackage{amsmath}
\usepackage{epsfig}
\usepackage{epstopdf}
\usepackage{bm}
\usepackage{graphicx,epsfig}
\usepackage{mathrsfs}
\usepackage{dcolumn}
\usepackage{color}
\usepackage{natbib}
\usepackage{CJK}
\def\be{\begin{equation}}
\def\ee{\end{equation}}
\def\bea{\begin{eqnarray}}
\def\eea{\end{eqnarray}}

\allowdisplaybreaks[2]

\begin{document}

\title{Sideband generation of transient lasing without population inversion}
\author{Luqi Yuan}
\affiliation{Texas A$\&$M University, College Station, TX 77843,
USA}
\author{Da-Wei Wang} \affiliation{Texas A$\&$M University,
College Station, TX 77843, USA}
\author{Christopher O'Brien}
\affiliation{Texas A$\&$M University, College Station, TX 77843,
USA}
\author{Anatoly A. Svidzinsky}
\affiliation{Texas A$\&$M University, College Station, TX 77843,
USA}
\author{Marlan O. Scully}
\affiliation{Texas A$\&$M University, College Station, TX 77843,
USA} \affiliation{Princeton University, Princeton, NJ 08544, USA}
\affiliation{Baylor University, Waco, TX 76798, USA}

\date{\today }

\begin{abstract}
We suggest a method to generate coherent short pulses by
generating a frequency comb using lasing without inversion in the
transient regime. We use a universal method to study the
propagation of a pulse in various spectral regions through an
active medium that is strongly driven on a low-frequency
transition on a time scale shorter than the decoherence time. The
results show gain on the sidebands at different modes can be
produced even if there is no initial population inversion
prepared. Besides the production of ultra-short pulse this
frequency comb may have applications towards making
short-wavelength or Tera-hertz lasers.
\end{abstract}


\maketitle

\emph{Introduction.} --- As a fundamental aspect of nonlinear
optics, optical sidebands generated via frequency modulation,
attracts widespread interest and have versatile applications in
atomic systems \cite{radeonychev,akulshin,knp}, terahertz quantum
cascade lasers \cite{cavalie}, ultrafast driven optomechanical
systems \cite{xiong}, polymer waveguides on a printed circuit
\cite{lanin}, and so on \cite{schliesser,goda}. In these cases, a
probe laser source is needed and optical sidebands are produced by
the interaction between the probe laser and the modulated medium.
However, the addition of this extra probe laser not only increases
the complexity of the experiment, but also introduces a limit to
this technology because it is difficult to prepare a table-top
laser pulse in some specific frequency regimes, \textit{i.e.} in
the extreme ultraviolet (XUV) or x-ray regime and the THz regime.

Lasing without inversion (LWI) \cite{olga,harris,scully89,
svidzinsky} has been studied in various media, such as in gas
\cite{fry}, circuit quantum electrodynamics \cite{marthaler}, and
terahertz intersubband-based devices \cite{pereira}. By preparing
an atomic system in a coherent superposition of states it is
possible to create atomic coherence to suppress absorption
resulting in LWI \cite{scullybook}. Steady-state LWI requires that
the spontaneous decay rate of the pumping transition is larger
than the decay rate of the lasing transition \cite{Imamoglu1991},
which is difficult to achieve when the frequency of the lasing
transition is higher than the drive field frequency. Thanks to a
recent experiment showing that a large collective atomic coherence
can be built up during a superradiant time scale much shorter than
the collisional decoherence time \cite{hui}, these obstacles can
be overcome in LWI in the transient regime \cite{svidzinsky},
where the lasing happens at a much shorter time than the
decoherence time and therefore all the decay rates can be
neglected. This paves the way for more complicated manipulation of
the quantum coherence to achieve sideband lasing at multiple
frequencies without initial population inversion.

In this letter, we combine the concepts of transient LWI and of
sideband generation to realize frequency comb generation at high
frequencies. The transient LWI is explored in a more complete
picture than any previous works by considering all the frequency
mode components. Through making a single-pass superradiant gain
\cite{siegman,qaser}, our results provide a new route toward
generating multiple-frequency coherent light and have implications
for the ultrashort pulse creation, short-wavelength coherent light
sources in the XUV and X-ray regime, and tunable THz laser
generation.

\begin{figure}[!h]
\includegraphics[width=6.5cm]{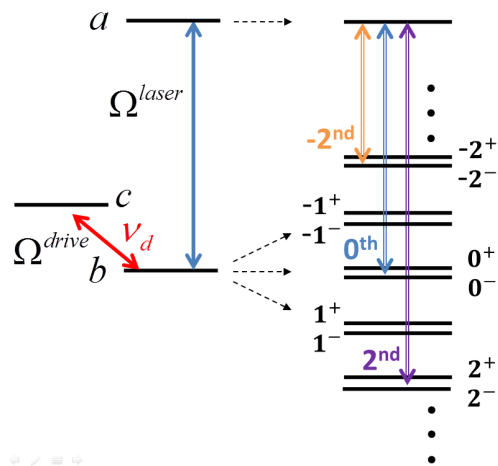}
\caption{Left: Energy diagram for V-scheme; Right: Floquet ladder
of states produced by the $c\rightarrow b$ transition driven by a
laser field with frequency $\nu_d$. Possible lasing transitions
are the $0^{\mathrm{th}}$-order transition ($\sim\omega_{ab}$),
and at the even sidebands $\pm2^{\mathrm{nd}}$-order
($\sim\omega_{ab}\pm2\nu_d$), etc. Each order is split in two
dressed states ($j^+$, $j^-$) by the rotating-wave terms of the
electric-dipole interaction. The side band signals are generated
due to the state mixing by the counter-rotating terms.}
\label{fig1}
\end{figure}

\emph{Model.} --- The mechanism of our proposal is shown in Fig.
\ref{fig1} (Left) based on a three-level V-type system. The system
is initially prepared such that most of the population remains in
the ground state but a little population is in the excited state
$|a\rangle$. A strong driving field $\Omega^{\text{drive}}$
propagates into the pencil-like active medium and couples the
transition $c \leftrightarrow b$. A Floquet ladder \cite{gibson}
is generated (see Fig. \ref{fig1} (Right)). The transitions from
$a$ to the Floquet ladder produce various lasing fields with
frequency $\nu_l\sim\omega_{ab} + 2j\nu_d$ ($j=0,\pm 1,\pm
2,\dots$) in a time scale much shorter than any decay time. Here
$\nu_l$ is the lasing frequency, $\omega_{ab}$ is the atomic
transition, $\nu_d$ is the driving field frequency. These fields
are coupled by $\Omega^{\text{drive}}$ via the atomic coherence.
The frequency difference between sidebands is always an even
multiple of $\nu_d$ since an atom in state $|i\rangle$ needs an
even number of photons to return to its original state
$|i\rangle$, through successive real and virtual processes (where
the counter-rotating terms play a role) \cite{picon,dawei}.

The simple but important physics behind the frequency comb gain
profile can be understood by considering the dressed-state
picture. By driving the $c\rightarrow b$ transition, the excited
state $a$ is coupled with two dressed states ($j^+$, $j^-$) at
each order in the Floquet ladder (see Fig. \ref{fig1}). Both of
the allowed transition frequencies are on the order of
$\omega_{ab} + 2j \nu_d$. The energy difference between two
dressed states depends on the drive field Rabi frequency
$\Omega_d$ and the drive field detuning $\Delta$. The initial
population in the ground state $b$ is redistributed to the two
split dressed states at each order. While there is no population
inversion in the bare-state system, it is still possible to
achieve transient lasing because of the population inversion in
dressed-state picture. Through the coupled atomic coherence,
different sideband modes are consequently amplified. The lasing
threshold can be reached by tuning the driving field intensity and
the medium length.

To start our analysis, we assume that $\rho_{bb}$, $\rho_{cc}$,
and $\rho_{cb}$ evolve only under the influence of the driving
field for the moment (see the corresponding equations in the
Appendix A) because the laser field coupled with $a
\leftrightarrow b$ transition is relatively weak. The drive field
$\Omega^{\mathrm{drive}} = \Omega_d \cos [\nu_d (t-z/c)]$ is
turned on adiabatically. We look for the solutions in the forms
$\rho_{bc} (t,z) = \sum_m \rho_{bc}^m e^{-im\nu_d(t-z/c)}$,
$\rho_{bb} (t,z) = \sum_m \rho_{bb}^m e^{-im\nu_d(t-z/c)}$. A set
of infinite coupled algebraic equations can be derived and the
solutions for $\rho_{bc}^m$ and $\rho_{bb}^m$ are found
numerically. The detail is in the Appendix B.

The propagation of the laser pulse is described by Maxwell's
equation \cite{scullybook}
\begin{equation}
\left( c^2 \frac{\partial^2}{\partial z^2} -
\frac{\partial^2}{\partial t^2} \right) \Omega^{\mathrm{laser}} =
\frac{2\Omega_a^2}{\omega_{ab}} \frac{\partial^2}{\partial t^2}
(\rho_{ab}+c.c.), \label{Eqmaxwell1}
\end{equation}
where $\Omega_a \equiv \sqrt{\frac{3N\lambda_{ab}^2\gamma
c}{8\pi}}$, where $N$ is the density, $\lambda_{ab}$ is the
$a\rightarrow b$ transition wavelength, $\gamma$ is the
$a\rightarrow b$ radiative decay rate, and $c$ is the speed of
light. The atomic coherences $\rho_{ab}$ and $\rho_{ac}$ evolve
with Eqs. (\ref{Eqrhoab1}) and (\ref{Eqrhoac1}) in the Appendix A.
We are looking for a solution in the form of a superposition of
spectral components without the rotating-wave-approximation (RWA)
\cite{radeonychev},
\begin{equation}
\Omega^{\mathrm{laser}} (t,z) = \sum_m \Omega_l^m (z)
e^{-i(\omega_{ab} + m \nu_d + \Delta\nu)(t-z/c)}+c.c. ,
\label{Deflaser}
\end{equation}
\begin{equation}
\rho_{ab} (t,z) = \sum_m \rho_{ab}^m (z) e^{-i(\omega_{ab} + m
\nu_d + \Delta\nu)(t-z/c)} , \label{Defrhoab}
\end{equation}
\begin{equation}
\rho_{ac} (t,z) = \sum_m \rho_{ac}^m (z) e^{-i(\omega_{ab} + m
\nu_d + \Delta\nu)(t-z/c)} , \label{Defrhoac}
\end{equation}
where $m=0, \pm 1, \pm2, \ldots$, and $\Delta\nu$ is the small
detuning of the lasing frequency from the frequency
$\omega_{ab}+m\nu_d$. By using the expressions in Eqs.
(\ref{Deflaser})-(\ref{Defrhoac}) and taking the components for
the same frequency mode $m$ with slowly-varying-envelope
approximation (SVEA), the equations of the evolution of the laser
field becomes
\begin{equation}
\frac{\partial}{\partial z} \Omega_l^m =
i\frac{\omega_m}{\omega_{ab}} \frac{\Omega_a^2}{c} \rho_{ab}^m,
\label{Eqmaxwell2}
\end{equation}
where $\omega_m \equiv \omega_{ab}+m\nu_d +\Delta\nu$. Here
introduce next set of coupled algebraic equations which combine
the equations that describe the evolution of the coherence
$\rho_{ab}$ and $\rho_{ac}$
\begin{equation}
\Phi_m^- \rho_{ab}^{m-2} + \Phi_m^0 \rho_{ab}^{m} + \Phi_m^+
\rho_{ab}^{m+2} = - \sum_q \Theta_m^{2q} \Omega_l^{m-2q},
\label{Eqrhoab3}
\end{equation}
where we define $\eta_m^\pm \equiv 1/\left( \omega_{cb} \pm \nu_d
+ m\nu_d + \Delta \nu +i\gamma_t \right)$, $\Phi_m^\pm \equiv
-\Omega_d^2 \eta_m^\pm /4$, $\Phi_m^0 \equiv \left(m\nu_d + \Delta
\nu +i\gamma_t \right) - \Omega_d^2 \left(\eta_m^- +\eta_m^+
\right)/4$, and $\Theta_m^{2q} \equiv
\rho_{bb}^{2q}-\rho_{aa}(0)\delta_{q0} + \Omega_d \left( \eta_m^-
\rho_{bc}^{2q-1} + \eta_m^+ \rho_{bc}^{2q+1} \right)/2$, where
$\gamma_t$ is the total decoherence rate, which is negligible in
the transient regime. Eq. (\ref{Eqrhoab3}) indicates that the
component of the field at the mode $m$ is coupled with those at
modes $m+2j$, where j is the integer.

We search for a solution of Eq. (\ref{Eqrhoab3}) with the form,
\begin{equation}
\Omega_l^m (z) = \sum_n u_n \varepsilon_n^m e^{i k_n z}.
\label{Omegalm}
\end{equation}
Using this form in Eq. (\ref{Eqmaxwell2}), we obtain $\rho_{ab}^m
(z) = \frac{\omega_{ab}}{\omega_{ab}+m\nu_d+\Delta\nu}
\frac{c}{\Omega_a^2}\sum_n u_n \varepsilon_n^m k_n e^{i k_n z}$.
With the trial solutions of $\Omega_l^m$ and $\rho_{ab}^m$, Eq.
(\ref{Eqrhoab3}) results in an infinite set of linear equations
with eigenvalues $k_n$ and their corresponding eigenvectors $\hat
\varepsilon_n = \left(\dots, \varepsilon_n^{m-2},
\varepsilon_n^{m}, \varepsilon_n^{m+2}, \dots\right)^T$. The
coefficient $u_n$ is determined by the boundary conditions for
$\Omega_l^m (z=0)$ and it reads $u_n = \sum_m \varepsilon_n^m
\Omega_l^m (z=0)$. There are an infinite coupled number of
frequency modes. However, the spectra must have a central spectral
region where all the frequency modes have relatively strong
intensities while the other frequencies far away from this region
fade out gradually. Therefore, we can solve Eq. (\ref{Eqrhoab3})
numerically in a central spectral region where it has central mode
$m = 0$ and boundary modes $m = m_0$. The set of infinite
equations is truncated to dimension $(m_0 + 1) \times (m_0 + 1)$
\cite{picon}.

\begin{figure}[t]
\includegraphics[width=8.0cm]{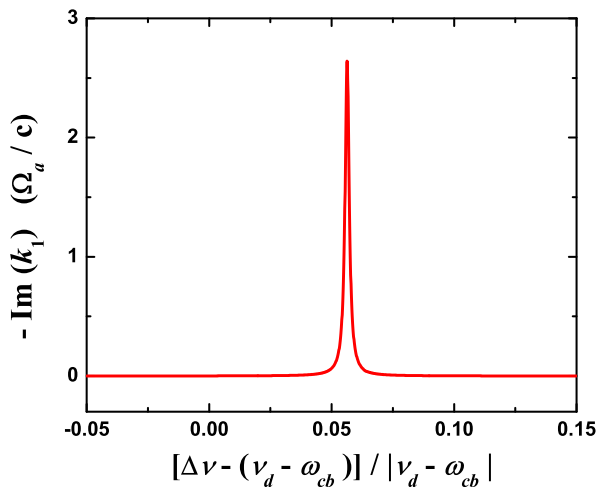}
\caption{The imaginary part of $k_1$ as a function of the lasing
frequency detuning $\Delta\nu$. The populations are $\rho_{aa}(0)
= 0.1$, $\rho_{bb}(0) = 0.9$ and $\rho_{cc}(0) = 0$, i.e., without
inversion. $\omega_{ab} = 5.0 \omega_{cb}$, $\Omega_a = 0.05
\omega_{cb}$, $\gamma_t = 10^{-4} \omega_{cb}$. We drive the
$c\rightarrow b$ transition with a weak detuned field with $\nu_d
= 1.1 \omega_{cb}$ and $\Omega_d = 0.05 \omega_{cb}$. We cut off
our calculation at $m=\pm 10$.} \label{eigenvalue}
\end{figure}

We first show the basic result in Fig. \ref{eigenvalue}. The gain
is characterized by the imaginary parts of eigenvalues $k_n$
($n=$1, 2,... with descending magnitudes of their imaginary parts)
of Eq. (\ref{Eqrhoab3}), since the fields generally follow $\sim
e^{-\text{Im}k_1 z}$. Especially, we focus on the leading
eigenvalue $k_1$ whose imaginary part has a magnitude several
orders larger than the rest. A peak of $-\mathrm{Im}(k_1)$ appears
at $\Delta\nu \sim 1.05 \Delta$ with width $\sim 0.01 \Delta$
where $\Delta \equiv\nu_d -\omega_{cb}$. We therefore can observe
sideband LWI in this region.

\begin{figure}
  \centering
  \begin{minipage}[c]{0.26\textwidth}
    \centering
    \includegraphics[width=7cm]{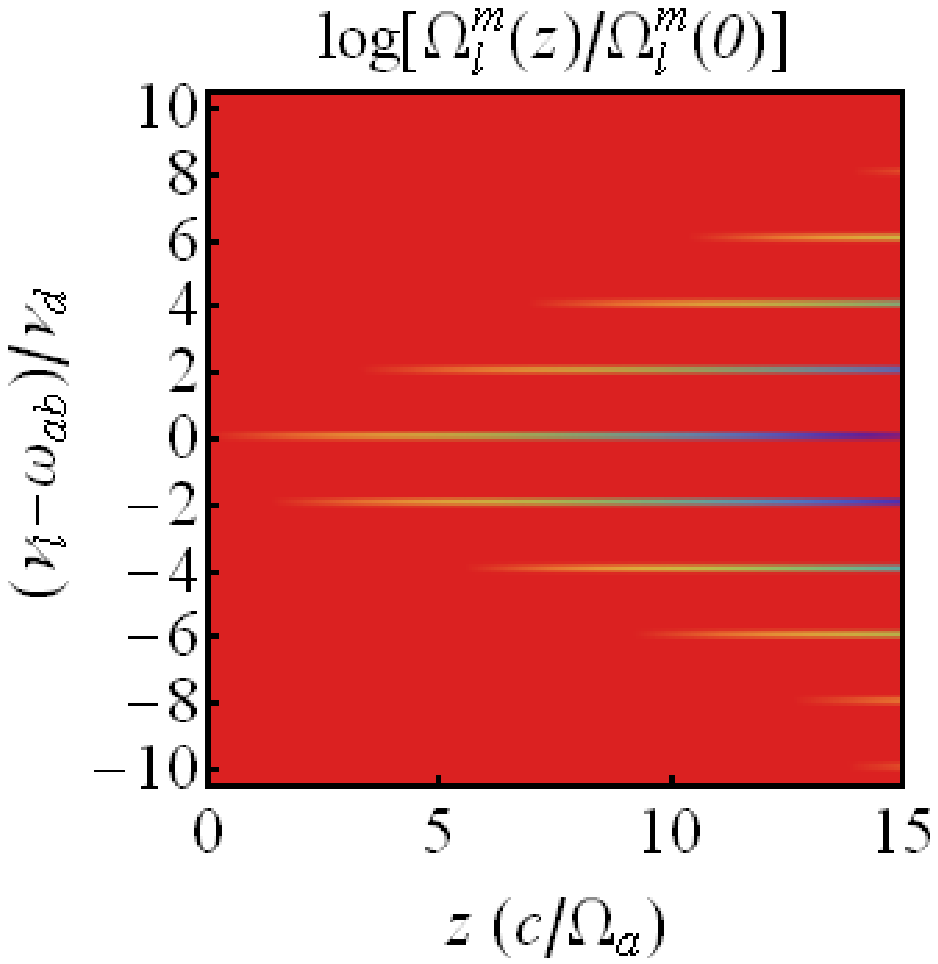}
  \end{minipage}%
  \begin{minipage}[c]{0.26\textwidth}
    \centering
    \includegraphics[width=0.8cm]{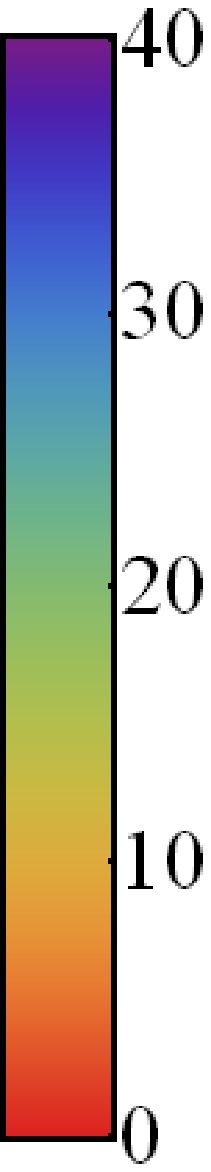}
  \end{minipage}
  \caption{The amplification of the output field in the whole spectral region with different propagation distance
  $z$. The creation of a frequency comb is shown. The plot is made with the same parameters as in Fig. \ref{eigenvalue}. } \label{gainplot}
\end{figure}

The amplitude of the output field at frequency mode $m$
($\Omega_l^m$) is determined by Eq. (\ref{Omegalm}). The gain of
each frequency component is not only dependent on the imaginary
part of the eigenvalues, but also dependent on the coefficients
such as $\varepsilon_n^m$, the elements in the eigenstates and
$u_n$ due to the boundary condition. It results in different
lasing amplifications for different frequency modes. If the field
component has smaller coefficients, it requires a longer
propagation length to be amplified. The result is plotted in Fig.
\ref{gainplot}. We find that we generate a frequency comb at a
long propagation distance ($z=15$ ($c/\Omega_a$)). With longer
propagation length, sideband lasing at the higher-order modes gets
amplified. For the field at mode $m \neq 0$ ($\Omega_l^m(z)$) with
frequency $\sim\omega_{ab}+m\nu_d$, the component of $k_1$ in Eq.
(\ref{Omegalm}) does not dominate over the components of the other
eigenvalues for small $z$, so the field component $\Omega_l^m(z)$
is not amplified compared to its initial value ($\Omega_l^m(0)$).
This means that the laser field has threshold behavior and the one
at a larger frequency mode has a higher threshold value (see Fig.
\ref{gainplot}). The amplification quantity
$\log[\Omega_l^m(L)/\Omega_l^m(0)]$ is linearly dependent on the
propagation length $L$ only if the propagation length $L$ exceeds
the threshold value. In this regime, the linear coefficients for
each curve at different frequency modes are the same because the
leading terms in Eq. (\ref{Omegalm}) for all modes $m$ are the
components of $k_1$ for large $z$ and all those terms grow
according to $\exp(-\mathrm{Im}k_1 z)$.

\begin{figure}[t]
\includegraphics[width=8.0cm]{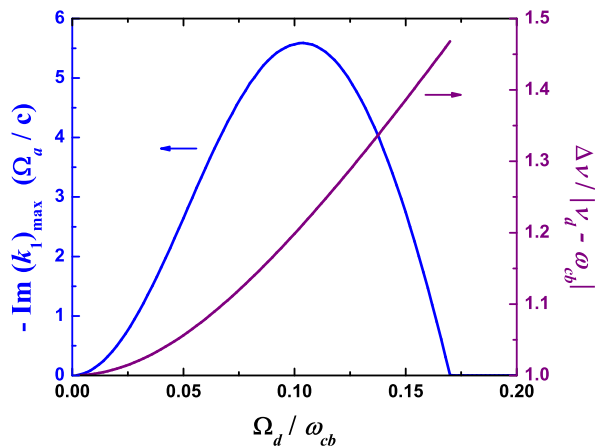}
\caption{The maximum value of the negative imaginary part of the
eigenvalue $k_1$, $-\mathrm{Im}(k_1)_{\mathrm{max}}$, (left blue)
with its corresponding lasing frequency detuning $\Delta\nu$
(right purple) for various drive field Rabi frequency $\Omega_d$.
The corresponding $\Delta\nu$ is only plotted for positive
$-\mathrm{Im}(k_1)$. This plot determines the Rabi frequency
$\Omega_d$ that should be chosen for peak gain.} \label{scaneigen}
\end{figure}

The amplification of the laser pulse in the whole spectral region
has a common source $-\mathrm{Im}(k_1)$. We study the relation
between $k_1$ and the drive field Rabi frequency $\Omega_d$ with
all the other parameters fixed. We solve Eq. (\ref{Eqrhoab3})
numerically for various $\Omega_d$ and search the maximum value of
$-\mathrm{Im}(k_1)_{\mathrm{max}}$, by scanning the lasing
frequency detuning $\Delta\nu$ for each set of parameters. The
dependence of the quantity $-\mathrm{Im}(k_1)_{\mathrm{max}}$ with
its corresponding lasing frequency detuning $\Delta\nu$ on
different $\Omega_d$ are plotted in Fig. \ref{scaneigen}. All of
the other parameters are the same as in Fig. \ref{eigenvalue}. We
find that the quantity $-\mathrm{Im}(k_1)_{\mathrm{max}}$ is
increasing with the drive field Rabi frequency $\Omega_d$ when
$\Omega_d$ is small. Nevertheless
$-\mathrm{Im}(k_1)_{\mathrm{max}}$ has a maximum after which it
drops counter-intuitively with increasing $\Omega_d$.

This behavior can be explained in the dressed-state picture. Each
order of the Floquet ladder of states is split in two dressed
states (as shown in Fig. \ref{fig1}). If $\Omega_d \rightarrow 0$,
one of the two dressed states (level $2j^-$ in the current case)
has $\sim 0$ population, but the corresponding coupling strength
between this dressed state and the excited state also goes to
zero. The increase of $\Omega_d$ leads to the enhancement of this
coupling strength and results in the increase of the gain.
However, larger $\Omega_d$ also leads to more population in this
dressed state, resulting in less population inversion. The
competition between these two mechanisms is the reason that the
quantity $-\mathrm{Im}(k_1)_{\mathrm{max}}$ has the maximum
positive value when $\Omega_d$ is near the resonance $\sim 0.1
\omega_{cb}$ (see Fig. \ref{scaneigen}). Only one of the two split
dressed states ($2j^-$) can have less population than the excited
state, so there is only one peak of the imaginary part of the
eigenvalue $k_1$ as shown in Fig. \ref{eigenvalue}. (This is
summarized in the detailed derivation in the Appendix C.) On the
other hand, changing the drive field Rabi frequency modifies the
energies of the two dressed states, so the corresponding lasing
frequency detuning $\Delta\nu$ is increasing versus $\Omega_d$
(right purple curve in Fig. \ref{scaneigen}).

\begin{figure}[t]
\includegraphics[width=7.0cm]{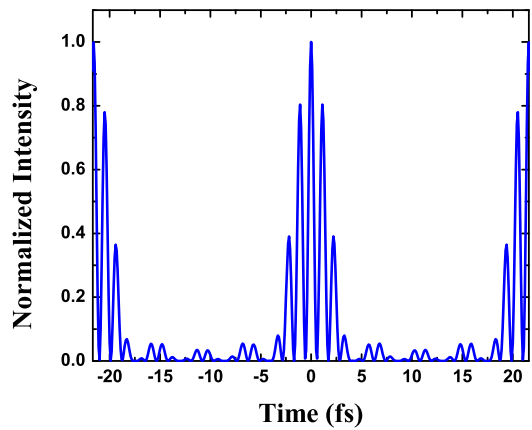}
\caption{Ultrashort pulse intensity in the case of Hydrogen
molecule. A 5-fs pulse is created with parameters given in the
text.}\label{shortpulse}
\end{figure}

The generated frequency comb has many applications:

\emph{Ultrashort pulse generation.} --- Ultrashort pulse
production can be achieved by modifying an input single-frequency
field to create an output field with multi-frequencies at the same
phase \cite{radeonychev,sokolov,zhi}. In contrast our method is
valid for generating the ultrashort pulse without the requirement
of an input field at the same centered frequency as the desired
output pulse. A low-frequency drive is used to modulate the
system. We choose Hydrogen molecule as an example, which has its
first vibrational transition frequency at the ground electronic
state $\lambda_{cb} = 2.28$ $\mu$m and a high-frequency electronic
transition ($B$ $^1\Sigma_u^+$ $\leftrightarrow$ $X$
$^1\Sigma_g^+$) at the frequency $\lambda_{ab} = 109$ nm.
Few-cycle pulse with 5 fs linewidth and 22 fs repetition period is
produced by converting the central 5 sideband LWIs at different
spectral components attenuated to equal values with
frequency-resolved filters (see Fig. \ref{shortpulse}). The
physical mechanism is similar to Ref. \cite{radeonychev}. By
changing to a different active medium, it is possible to further
shorten the pulse width.

\emph{Short-wavelength laser.} --- The conversion from the
long-wavelength drive pulse to the short-wavelength emission
pulse, in particular the pulses at the blue-shifted sidebands
($m>0$), provides a promising choice for generating a
high-frequency laser.  Consider as a proof-of-principle, the
realistic experimental choice of a helium plasma gas which is
partially excited to the metastable triplet state, 2
$^3\mathrm{S}_1$, as realized in one recent experiment \cite{hui}.
Where the density is $\sim 10^{16}$ cm$^{-3}$ and we can drive the
infrared transition 2 $^3\mathrm{P}_1\rightarrow$2
$^3\mathrm{S}_1$ (1083 nm) with a drive field wavelength as
$\lambda_d=1022$ nm and a Rabi frequency of $\Omega_d \sim
10^{14}$ rad/s. The dispersion of the drive field is negligible if
it is detuned significantly from the resonance. A little
population is left in the excited state 3 $^3\mathrm{P}_1$ by
non-radiative three-body recombination following an optical field
ionization. This allows transient LWI to occur at the ultraviolet
transition 3 $^3\mathrm{P}_1\rightarrow$2 $^3\mathrm{S}_1$ (388.9
nm). The higher-order sideband lasing would have wavelengths as
$\lambda_l^{(2)} \sim 220.8$ nm, $\lambda_l^{(4)} \sim 154.2$ nm,
etc. In a 1-cm long medium, we find a single-pass nano-Joule level
coherent emission at the wavelength $\sim 220.8$ nm with the
parameters listed above. In principle, this method can make
table-top laser pulses in XUV and X-ray regime with a visible
driving field.

\emph{Tunable THz laser.} --- Graphene has suitable energy level
structure with strong dipole moments to study the physics in the
THz regime in a magnetic field \cite{yao,tokman}. The V-scheme
model is composed of the Landau levels (LLs) near the Dirac point
with energy quantum numbers -2, -1, and 3. $B \sim 1.4$ mT gives
the transition frequency $\omega_{cb} / 2 \pi \sim 140$ GHz
between LLs with energy quantum numbers -2 and -1, which can be
driven by commercially available facilities for coherent
millimeter wave source. Propagation effects are complicated by the
large number of graphene layers \cite{jung}; our model shows
sideband LWI at frequencies $1.06 \pm 0.28 n$ THz ($n =
0,1,2,\ldots$). The transition frequencies in the V-scheme can be
further changed by modifying the magnetic field. The emission at
different frequencies in THz regime can be used to build a tunable
THz laser.

In conclusion, we study Frequency comb generation via sideband
transient LWI. We use the Floquet method to solve the system in
the weak lasing field limit (the population is unchanged due to
the lasing field) and find amplified emission at different
frequency modes. Threshold behavior is seen for high-order
sidebands. This universal model has many possible applications
including ultrashort pulse generation, short-wavelength laser in
the XUV and X-ray regime, and tunable THz laser source. We gave an
example scheme for the generation of 5-fs pulses in molecular
hydrogen.

\begin{acknowledgments}
The authors thank O. Kocharovskaya for useful discussion. We
acknowledge the support of the National Science Foundation Grants
PHY-1241032, PHY-1205868 and the Robert A. Welch Foundation
(Awards A-1261). L.Y. is supported by the Herman F. Heep and
Minnie Belle Heep Texas A$\&$M University Endowed Fund
held/administered by the Texas A$\&$M Foundation.
\end{acknowledgments}

\appendix

\section*{Appendix A: Density matrix equations for a V-scheme model}

\renewcommand{\theequation}{A-\arabic{equation}}
\setcounter{equation}{0}

Here we list the full-set of the density matrix equations for a
V-scheme model shown in Fig. 1,
\begin{equation}
\dot \rho_{ab} = -(i \omega_{ab}+\gamma_t) \rho_{ab} +
i\Omega^{\mathrm{laser}} (\rho_{bb} - \rho_{aa}) -
i\Omega^{\mathrm{drive}} \rho_{ac}, \label{Eqrhoab1}
\end{equation}
\begin{equation}
\dot \rho_{ac} = -(i\omega_{ac}+\gamma_t) \rho_{ac} +
i\Omega^{\mathrm{laser}} \rho_{bc} - i\Omega^{\mathrm{drive}*}
\rho_{ab}, \label{Eqrhoac1}
\end{equation}
\begin{equation}
\dot \rho_{cb} = -(i\omega_{cb}+\gamma_t) \rho_{cb} +
i\Omega^{\mathrm{drive}} (\rho_{bb} - \rho_{cc}) -
i\Omega^{\mathrm{laser}} \rho_{ca}, \label{Eqrhocb}
\end{equation}
\begin{equation}
\dot \rho_{bb} = - i\Omega^{\mathrm{drive}} \rho_{bc} +
i\Omega^{\mathrm{drive}*} \rho_{cb} - i\Omega^{\mathrm{laser}}
\rho_{ba} + i\Omega^{\mathrm{laser}*} \rho_{ab}, \label{Eqrhobb}
\end{equation}
\begin{equation}
\dot \rho_{cc} = i\Omega^{\mathrm{drive}} \rho_{bc} -
i\Omega^{\mathrm{drive}*} \rho_{cb} ,
\end{equation}
\begin{equation}
\rho_{aa} + \rho_{bb} + \rho_{cc} = 1,
\end{equation}
where $\gamma_t$ is the total decoherence rate. These equations
are supplemented by Maxwell's equation
\begin{equation}
\left( \frac{\partial^2}{\partial z^2} - \frac{1}{c^2}
\frac{\partial^2}{\partial t^2} \right) E^{\mathrm{laser}} = \mu_0
\frac{\partial^2 P^{\mathrm{laser}}}{\partial t^2},
\end{equation}
where $P^{laser} = N(\wp_{ba}\rho_{ab}+c.c.)$.

\section*{Appendix B: Floquet equations for the two-level system with a detuned drive field}

\renewcommand{\theequation}{B-\arabic{equation}}
\setcounter{equation}{0}

Here we consider two-level system ($c\rightarrow b$) with a
detuned drive field $\Omega^{\mathrm{drive}} = \Omega_d \cos[\nu_d
(t-z/c)]$ as shown in Fig. 1. We look for the solutions in the
forms $\rho_{bc} (t,z) = \sum_m \rho_{bc}^m e^{-im\nu_d(t-z/c)}$
and $\rho_{bb} (t,z) = \sum_m \rho_{bb}^m e^{-im\nu_d(t-z/c)}$ for
the equations,
\begin{equation}
\dot \rho_{bc} = (i\omega_{cb} - \gamma/2) \rho_{bc} -
i\Omega^{\mathrm{drive}} \left( \rho_{bb} - \rho_{cc}\right),
\end{equation}
\begin{equation}
\dot \rho_{bb} = \gamma \rho_{cc} -
i\Omega^{\mathrm{drive}}\rho_{bc} +
i\Omega^{\mathrm{drive}*}\rho_{cb},
\end{equation}
\begin{equation}
\rho_{bb} + \rho_{cc} = \rho_{bb}(0) + \rho_{cc}(0),
\end{equation}
where the depopulation decay rate $\gamma$ is very small compared
with all other parameters. Therefore, the set of coupled algebraic
equations are found to be
\begin{equation*}
\left(m\nu_d+\omega_{cb}+i\gamma/2\right)\rho_{bc}^m - \Omega_d
\left(\rho_{bb}^{m-1}+\rho_{bb}^{m+1}\right)
\end{equation*}
\begin{equation}
= -\frac{\Omega_d}{2} (\delta_{m,1}+\delta_{m,-1})
[\rho_{bb}(0)+\rho_{cc}(0)], \label{EqApprhobc}
\end{equation}
\begin{equation*}
(m\nu_d+i\gamma)\rho_{bb}^m - \frac{\Omega_d}{2}
\left(\rho_{bc}^{m+1}+\rho_{bc}^{m-1}-\rho_{bc}^{-m+1*}-\rho_{bc}^{-m-1*}\right)
\end{equation*}
\begin{equation}
= i\gamma[\rho_{bb}(0)+\rho_{cc}(0)]\delta_{m0}.
\label{EqApprhobb}
\end{equation}
General results for $\rho_{bc}^m$ and $\rho_{bb}^m$ can be found
by solving infinite coupled Eqs. (\ref{EqApprhobc}) and
(\ref{EqApprhobb}) numerically. Note from Eq. (\ref{EqApprhobb})
that $\rho_{bb}^m = \rho_{bb}^{-m*}$, which leads to the real
solution for $\rho_{bb}$.

\section*{Appendix C: LWI in the dressed state picture}

\renewcommand{\theequation}{C-\arabic{equation}}
\setcounter{equation}{0}

Here we consider only the $0^{\mathrm{th}}$-order lasing
transition in the dressed state picture. The drive field couples
the $c\rightarrow b$ transition and has the form
$\Omega^{\mathrm{drive}} = \Omega_d \cos (\nu_d \tau)$, where
$\tau = t-z/c$. With the rotating-wave-approximation, the
interaction Hamiltonian is
\begin{equation}
V = - \hbar \Delta |c\rangle\langle c| - \frac{\hbar \Omega_d}{2}
|c\rangle\langle b| - \frac{\hbar \Omega_d}{2} |b\rangle\langle
c|,
\end{equation}
where $\Delta = \nu_d - \omega_{cb}$. It has two eigenstates as
\begin{equation}
|+\rangle =
\sqrt{\frac{\Omega_{\mathrm{eff}}-\Delta}{\Omega_{\mathrm{eff}}}}|c\rangle
-
\sqrt{\frac{\Omega_d^2}{2\Omega_{\mathrm{eff}}(\Omega_{\mathrm{eff}}-\Delta)}}|b\rangle,
\end{equation}
\begin{equation}
|-\rangle =
\sqrt{\frac{\Omega_{\mathrm{eff}}+\Delta}{\Omega_{\mathrm{eff}}}}|c\rangle
+
\sqrt{\frac{\Omega_d^2}{2\Omega_{\mathrm{eff}}(\Omega_{\mathrm{eff}}+\Delta)}}|b\rangle,
\end{equation}
where $\Omega_{\mathrm{eff}} \equiv \sqrt{\Omega_d^2 + \Delta^2}$
and their corresponding eigenvalues are
\begin{equation}
\omega_{\pm} = \frac{1}{2} (-\Delta \pm \Omega_{\mathrm{eff}}).
\end{equation}
For a system which is initially at state $|b\rangle$ at $\tau =
0$, the system evolves as
\begin{equation*}
|\psi(\tau)\rangle = -
\sqrt{\frac{\Omega_{\mathrm{eff}}+\Delta}{2\Omega_{\mathrm{eff}}}}
\sqrt{\rho_{bb}(0)}e^{-i\omega_{+}\tau}|+\rangle
\end{equation*}
\begin{equation}
+\sqrt{\frac{\Omega_{\mathrm{eff}}-\Delta}{2\Omega_{\mathrm{eff}}}}
\sqrt{\rho_{bb}(0)}e^{-i\omega_{-}\tau}|-\rangle
\end{equation}
at $\tau = t-z/c \geq 0$. Therefore, the density matrix elements
are
\begin{equation}
\rho_{++}(t,z) =
\frac{\Omega_{\mathrm{eff}}+\Delta}{2\Omega_{\mathrm{eff}}}
\rho_{bb}(0),
\end{equation}
\begin{equation}
\rho_{--}(t,z) =
\frac{\Omega_{\mathrm{eff}}-\Delta}{2\Omega_{\mathrm{eff}}}
\rho_{bb}(0),
\end{equation}
\begin{equation}
\rho_{+-}(t,z) =
-\frac{\sqrt{\Omega_{\mathrm{eff}}^2-\Delta^2}}{2\Omega_{\mathrm{eff}}}
\rho_{bb}(0)e^{-i(\omega_+-\omega_-)(t-z/c)},
\end{equation}
\begin{equation}
\rho_{-+}(t,z) =
-\frac{\sqrt{\Omega_{\mathrm{eff}}^2-\Delta^2}}{2\Omega_{\mathrm{eff}}}
\rho_{bb}(0)e^{i(\omega_+-\omega_-)(t-z/c)},
\end{equation}

Now, we introduce weak lasing field $E_l$ with frequency $\nu_l
\sim \omega_{ab}$ coupling the $a\rightarrow b$ transition. The
Hamiltonian reads
\begin{equation*}
H = \omega_a |a\rangle\langle a| + \omega_+ |+\rangle\langle +| +
\omega_- |-\rangle\langle -|
\end{equation*}
\begin{equation*}
- \wp_{ab}E_le^{-i\nu_lt}|a\rangle\langle b| -
\wp_{ba}E_l^*e^{i\nu_lt}|b\rangle\langle a|
\end{equation*}
\begin{equation*}
= \omega_a |a\rangle\langle a| + \omega_+ |+\rangle\langle +| +
\omega_- |-\rangle\langle -|
\end{equation*}
\begin{equation}
+ \left(-\wp_{a+}E_le^{-i\nu_lt}|a\rangle\langle +| -
\wp_{a-}E_le^{-i\nu_lt}|b\rangle\langle -| + H.c.\right),
\end{equation}
where
\begin{equation}
\wp_{a+} \equiv
-\sqrt{\frac{\Omega_{\mathrm{eff}}+\Delta}{2\Omega_{\mathrm{eff}}}}\wp_{ab},
\end{equation}
\begin{equation}
\wp_{a-} \equiv
\sqrt{\frac{\Omega_{\mathrm{eff}}-\Delta}{2\Omega_{\mathrm{eff}}}}\wp_{ab}.
\end{equation}
We assume that $E_l$ is so weak that it doesn't change the
populations and the coherence between states $|+\rangle$ and
$|-\rangle$. Therefore we find
\begin{equation*}
\frac{d}{dt} \tilde\rho_{a+} = -i(\omega_{a+} - \nu_l)
\tilde\rho_{a+}
\end{equation*}
\begin{equation}
- i\wp_{a+}E_l\left[\rho_{aa}(0)-\rho_{++}\right] +
i\wp_{a-}E_l\rho_{-+},
\end{equation}
\begin{equation*}
\frac{d}{dt} \tilde\rho_{a-} = -i(\omega_{a-} - \nu_l)
\tilde\rho_{a-}
\end{equation*}
\begin{equation}
- i\wp_{a-}E_l\left[\rho_{aa}(0)-\rho_{--}\right]
+ i\wp_{a+}E_l\rho_{+-},
\end{equation}
where $\omega_{a\pm} \equiv \omega_a - \omega_\pm$, and $\tilde
\rho_{a\pm} \equiv \rho_{a\pm} e^{i\nu_l t}$. The Maxwell's
equation has the expression
\begin{equation}
\left(\frac{\partial}{\partial t} + c \frac{\partial}{\partial
z}\right) E_l = \frac{i\nu_l}{2\epsilon_0} \wp_{ab}\tilde\rho_{ab}
=  \frac{i\nu_l}{2\epsilon_0}
\left(\wp_{a+}\tilde\rho_{a+}+\wp_{a-}\tilde\rho_{a-}\right).
\end{equation}
If we take RWA and neglect all the fast-oscillating terms, then
$E_l$ is only possible to get amplified at the resonant frequency
$\nu_l^{\pm} = \omega_{a\pm}$ with the corresponding coherence as
\begin{equation}
\frac{d}{dt} \tilde\rho_{a\pm} = -
i\wp_{a\pm}E_l\left[\rho_{aa}(0)-
\frac{\Omega_{\mathrm{eff}}\pm\Delta}{2\Omega_{\mathrm{eff}}}
\rho_{bb}(0)\right].
\end{equation}
Hence the electrical field $E_l$ evolves as
\begin{equation*}
\left(\frac{\partial}{\partial t} + c \frac{\partial}{\partial
z}\right) \dot E_l
\end{equation*}
\begin{equation}
=\frac{\nu_l \wp_{ab}^2}{2\epsilon_0} \left\{
\frac{\Omega_{\mathrm{eff}}\pm\Delta}{2\Omega_{\mathrm{eff}}}
\left[\rho_{aa}(0) -
\frac{\Omega_{\mathrm{eff}}\pm\Delta}{2\Omega_{\mathrm{eff}}}
\rho_{bb}(0)\right] \right\} E_l.
\end{equation}

From this result, we find that the electrical field $E_l$ can get
amplified if there is population inversion between state
$|a\rangle$ and state $|\pm\rangle$ in the dressed state picture.
We consider the case that $\rho_{aa}(0) \ll \rho_{bb}(0)$ and
assume $\Delta > 0$. Lasing happens at the transition between the
state $|a\rangle$ and the state $|-\rangle$ and gain is dependent
on the quantity
$\frac{\Omega_{\mathrm{eff}}-\Delta}{2\Omega_{\mathrm{eff}}}
\left[\rho_{aa}(0) -
\frac{\Omega_{\mathrm{eff}}-\Delta}{2\Omega_{\mathrm{eff}}}
\rho_{bb}(0)\right]$. When $\Omega_d\rightarrow 0$, gain
$\rightarrow 0$ since
$\frac{\Omega_{\mathrm{eff}}-\Delta}{2\Omega_{\mathrm{eff}}}\rightarrow
0$ though there is population inversion $\rho_{aa}(0) >
\frac{\Omega_{\mathrm{eff}}-\Delta}{2\Omega_{\mathrm{eff}}}
\rho_{bb}(0)$. Gain is increasing with the increase of $\Omega_d$
initially. After it reaches the maximum value, it will decrease
until it becomes zero when there is no population inversion
$\rho_{aa}(0) \le
\frac{\Omega_{\mathrm{eff}}-\Delta}{2\Omega_{\mathrm{eff}}}
\rho_{bb}(0)$ for a very large $\Omega_d$. The corresponding
lasing frequency is $\nu_l = \omega_{a-} = \omega_{ab} +
\frac{1}{2} (\Delta + \Omega_{\mathrm{eff}})$, which is increasing
with $\Omega_d$. It has the similar result for the case $\Delta <
0$ and the lasing happens at the transition between the state
$|a\rangle$ and the state $|+\rangle$.

\section*{Appendix D: Numerical simulation with the
full-set of Maxwell and Schr\"{o}dinger equations}

\renewcommand{\theequation}{D-\arabic{equation}}
\setcounter{equation}{0}

\begin{figure}[!h]
\includegraphics[width=7.5cm]{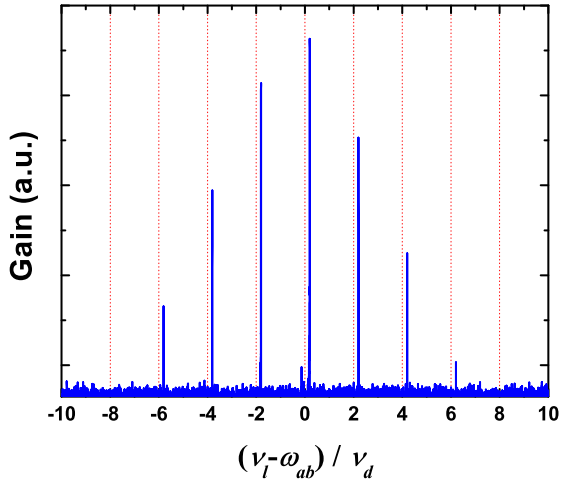}
\caption{Detailed numerical experiments with parameters: $\nu_d =
1.06 \omega_{cb}$, $\Omega_d = 0.18 \omega_{cb}$, $\Omega_a =
0.0754 \omega_{cb}$, $L = 7.54 c/\Omega_a$, $\rho_{aa}(0)=0.15$,
$\rho_{bb}(0)=0.85$, $\rho_{cc}(0)=0$, and $\gamma_t = 10^{-4}
\omega_{cb}$.} \label{LWIFFT}
\end{figure}

Finally, we show the detailed numerical simulation with the
full-set of Maxwell and Schr\"{o}dinger equations including
population evolutions without any approximation except SVEA in
Fig. \ref{LWIFFT}. We use the polarization source term in the
equations to describe production rate of the dipole due to the
spontaneous emission \cite{feld}. We see multiple single-pass gain
peaks above the noise level and they are located at the lasing
frequencies $\nu_l^{\pm 2n} \sim \omega_{ab} \pm 2 n \nu_d$. There
is no population inversion in the system. Coherent emission is
generated directly from vacuum fluctuations without an initial
seed pulse. The results of the amplification are generally
linearly dependent on $\Omega_a L$. This feature gives us
flexibility for choosing parameters in future experiments. For
example, if the system has a smaller $\Omega_a$ than what we
propose, it can still produce the same amount of gain as what we
expect by increasing $L$.

\end{document}